\documentclass[aps,%
12pt,%
final,%
oneside,%
onecolumn,%
nobibnotes,%
nofootinbib,%
superscriptaddress,%
centertags,
floatfix,
secnumarabic]%
{revtex4}%
\usepackage{graphicx}
\usepackage{bm}
\RequirePackage[utf8]{inputenc}
\RequirePackage[T2A]{fontenc}
\RequirePackage[russian,english]{babel}
\usepackage{amsmath}

\begin{document}

\selectlanguage{english}
\normalsize
\title{The production of excited states of doubly heavy baryons at the Large Hadron Collider. }
\author{\firstname{A.~V.}~\surname{Berezhnoy}}
\email{Alexander.Berezhnoy@cern.ch}
\affiliation{SINP MSU, Moscow, Russia}

\author{\firstname{I.~N.}~\surname{Belov}}
\email{in.belov@physics.msu.ru}
\affiliation{Physics department of MSU, Moscow, Russia}

\author{\firstname{A.~K.}~\surname{Likhoded}}
\email{Anatolii.Likhoded@ihep.ru}
\affiliation{NRC “Kurchatov Institute”–IHEP, Protvino, Russia}

\begin{abstract}
\normalsize
In the framework of diquark model for production we discuss the yields and observation prospects for excited states of doubly heavy baryons at LHC kinematic conditions. 
\end{abstract}

\maketitle

\section{INTRODUCTION}

Production and decay challenges for doubly heavy baryons attract researches for more than two decades (see, for example,~ \cite{Ebert:1996ec,Kiselev:2001fw}). Perhaps such an interest is caused by their exciting structure. Actually since such hadrons consist of two heavy quarks and one light quark, it is quite reasonable to divide them into two subsystems: compact doubly heavy diquark and a light quark. States of doubly heavy diquark (antitriplet by colour) can be described within the same models as quarkonia states (potential models for instance). Since the spectroscopy of quarkonia is described quite well under the decay threshold to open flavour, the diquark spectroscopy has a chance to be described adequately as well. Assuming that such a diquark is a compact, antitriplet by colour object one can consider its interaction with a light quark as an interaction between a quark and an antiquark. This approach greatly simplifies theoretical research of doubly heavy baryons and allows one to obtain detailed predictions for properties of such systems (see, for example,~\cite{Ebert:1996ec,Kiselev:2002an,Kiselev:2001fw,Faustov:2018vgl}).

It is worth to mention that the spectroscopy of doubly heavy baryons can be investigated not only in the framework of quark-diquark approach but also through direct solving the quantum three body problem (see., for instance,~\cite{Kerbikov:1987vx, Albertus:2006wb, Albertus:2006ya,Roncaglia:1995az,Roberts:2007ni,Yoshida:2015tia}). It is very significant direction of research but we should notice that currently it is not possible to unambiguously conclude that quantum problem of three bodies is more correct approximation than quark-diquark one. Thus lattice calculations testify to so-called ``Y''-shaped interation, from which the quark-diquark approximation arises naturally. One more argument for the latest approach is the fact that in case of spectroscopy of baryons with one heavy quark the model of interaction between heavy quark and a light diquark works well. In addition it is worth noting that Regge trajectories for light mesons and light baryons have the same slope which argues in favour of quark-diquark model in case of light hadrons.

While studying the spectroscopy of doubly heavy baryons one can choose between the two approaches mentioned here, studying the production of doubly heavy baryons doesn't provide this choice, and the only more or less consistent model of their production known today is based on the assumption that initially produced diquark transforms into doubly heavy baryon. It is clear that the production of a heavy diquark is very similar to associative production of quarkonium and a heavy quark: in both cases production of two pairs of heavy quarks and the consequent formation of doubly heavy system take place. But there is an essential difference. The point is as it is shown in works~\cite{Kom:2011bd,Baranov:2011ch,Berezhnoy:2012xq, Berezhnoy:2015jga}, associative production of quarkonium with hidden flavour and a heavy quark acquires significant contribution from so-called mechanism of doubly parton scattering (DPS), in which quarkonium and its attendant heavy quark are produced in different parton collisions. Unlike the heavy quarkonium production where DPS provides a comparable contribution in case of diquark production DPS mechanism is suppressed. Independent production of two pairs of heavy quarks doesn`t allow them to merge into diquark. That's why we expect the yields of doubly heavy baryons $\Xi_{cc}$ and $\Xi_{bb}$ to be much less than yields of associative production of the corresponding quarkonia and heavy quarks.

\begin{figure}
 \centering 
\resizebox*{0.7\textwidth}{!}{\includegraphics{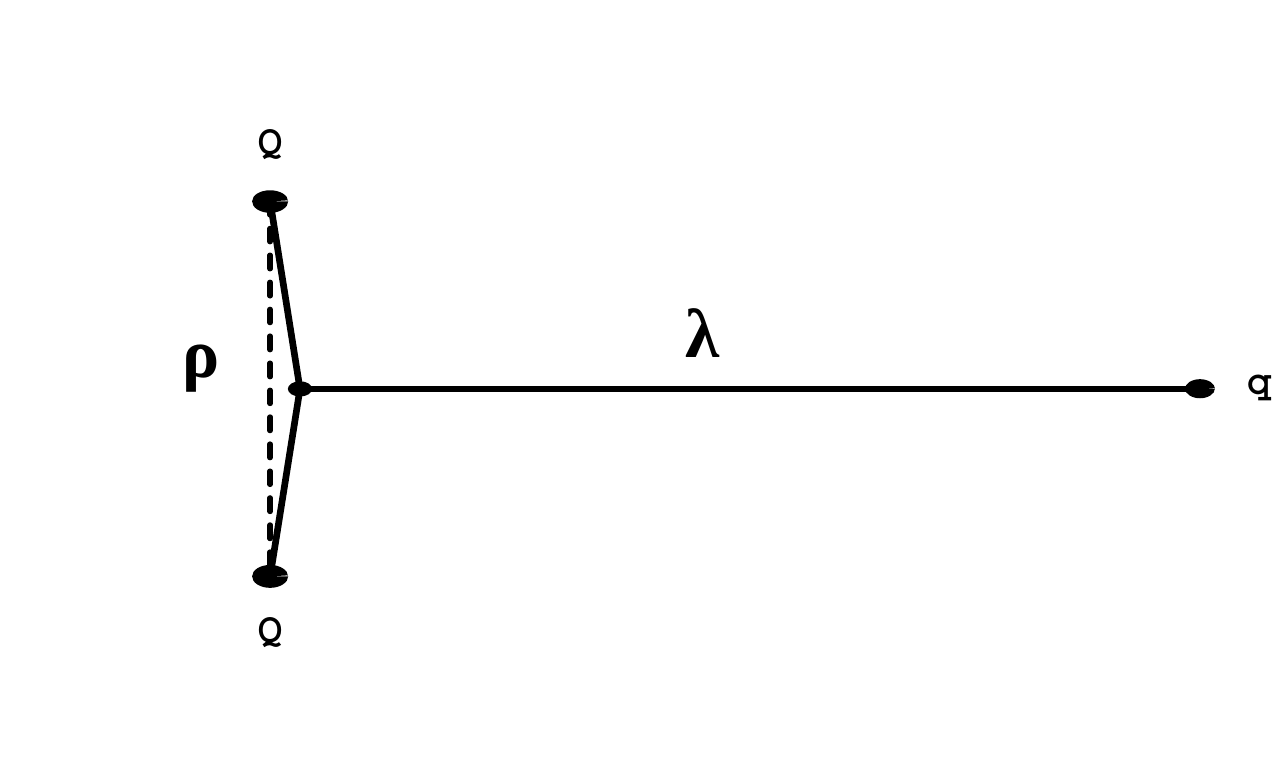}}
\caption{ Schematic representation of $\rho$ and $\lambda$ excited states of $\Xi_{QQ}$ baryon: $\rho$ states are states with excited diquark, $\lambda$ are states with excited light quark.}
\label{fig:excitations}
\end{figure}

While theoretical research of doubly heavy baryons is being carried out for many years, the first experimental observation of such state was achieved quite recently by collaboration LHCb: in 2017 year baryon with two charmed quarks $\Xi_{cc}^{++}$ 
was observed in decay mode $\Lambda_c^+ K^- \pi^+ \pi^+$ \cite{Aaij:2017ueg}. This observation is already confirmed in mode $\Xi_c^+ \pi^+$ \cite{Aaij:2018gfl}. The lifetime of this new state is also measured~\cite{Aaij:2018wzf}. In this work we discuss the perspectives for doubly heavy baryons further studying, in particular we estimate the yields of doubly heavy baryons with excited heavy diquark --- so-called $\rho$-excitations (see Fig.~\ref{fig:excitations}).

\section{Calculation technique}

Since the calculation technique is detailed in paper~\cite{Berezhnoy:2018krl}, in this work we provide its outline only. In the context of quark-diquark production model the baryon production is natural to be divided into two stages. At the first stage of calculations doubly heavy diquark in antitriplet colour state is produced perturbatively in hard interaction; at the second --- doubly heavy diquark turns into baryon in the soft hadronization process (see, for instance,~\cite{Berezhnoy:1995fy, Berezhnoy:1998aa, Baranov:1995rc, Baranov:1997sg, Chang:2006eu}). Typically the hadronization process is considered within the fragmentation approach by analogy with hadronization of one heavy quark into heavy hadron.   

As it is mentioned in work~\cite{Gershtein:2000nx}, the soft gluon emission complicates the classification of levels of heavy diquark with quarks of different flavours and therefore we will consider only $cc$- and $bb$-diquarks further.

Assuming slight dependence of production amplitude of four heavy quarks $T_{Q \bar Q Q \bar Q}$ on three-momentum $\boldsymbol{q}$ of a quark within the diquark, the diquark production amplitude can be expanded into a series over $\boldsymbol{q}$ powers: 
\begin{equation}
A \sim 
\int d^3\boldsymbol{q}\,\Psi^*_{[Q Q]_{\boldsymbol{3}}}(\boldsymbol{q})\left\{ \bigl. T_{Q \bar Q Q \bar Q} \bigr|_{\boldsymbol{q}=0} +
\bigl.\boldsymbol{q}\frac{\partial}{\partial \boldsymbol{q}} T_{Q \bar Q Q \bar Q} \bigr|_{\boldsymbol{q} =0}  +  \dots \right\},
\label{eq:q_expasion}
\end{equation}
where $\Psi_{[Q Q]_{\boldsymbol{3}}}(\boldsymbol{q})$ is a wave function of the diquark in colour antitriplet. The first term in the expansion (\ref{eq:q_expasion}) will provide the major contribution to production of the $S$-wave diquark, the second term --- to production of the $P$-wave one.

Requirement for antisymmetry for the wave function of a diquark with two identical quarks puts a restriction on its spin: $S$-wave diquark can only have spin~$1$ while $P$-wave diquark --- only spin $0$. Thus production amplitude of $S$-wave state of the diquark is determined by formula 
\begin{equation}
A^{s_z} = \frac{1}{\sqrt{4\pi}}R_S(0) \cdot \bigl. T_{Q\bar Q  Q\bar Q }^{s_z} \bigr|_{\boldsymbol{q}=0},
\end{equation}
where $s_z$ is a diquark spin projection and $R_S(0)$ is value of radial wave function at origin;
and production amplitude of $P$-wave state of the diquark is determined by formula
\begin{equation}
A^{l_z} = i\sqrt{\frac{3}{4\pi}} R_P'(0) \cdot \bigl.\{{\cal L}^{l_z} T_{Q \bar Q Q \bar Q}\}\bigr|_{\boldsymbol{q}=0}, 
\end{equation}
where $l_z$ is a diquark orbital momentum projection, $R_P'(0)$ --- derivative of radial wave function at origin and ${\cal L}^{l_z}$ --- differential operator of the following form:

\begin{equation}
{\cal L}^{l_z} = \begin{cases}
{\cal L}^{-1}=\frac{1}{\sqrt{2}}\left (\frac{\partial}{\partial q_x}
+i\frac{\partial}{\partial q_y} \right ) \\
{\cal L}^0=\frac{\partial}{\partial q_z} \\
{\cal L}^{+1}=-\frac{1}{\sqrt{2}}\left (\frac{\partial}{\partial q_x}
-i\frac{\partial}{\partial q_y} \right ).\\
\end{cases}
\end{equation}

The resulting colour antitriplet should be hadronized to form a baryon. Since light quark with effective mass $m_q$ in the baryon with mass~$M$ picks up approximately $\frac{m_q}{M}$ from the whole transverse momentum of the baryon, this quark is always present in a quark sea at LHCb kinematic conditions. That's why one can assume that doubly heavy baryon hadronizes by joining with one of the light quarks $u$, $d$ or $s$ in the same proportion $1:1:0.26$~~as $b$ quark~\cite{Aaij:2011jp}.
We also suggest that it hadronizes with unit probability. The latter assumption is pretty much a guess because diquark has a color charge and therefore strongly interacts with its environment, that could lead to the diquark dissociation.

On the other hand it can be assumed that diquark hadronizes according to fragmentation model in line with heavy meson. Under this model the diquark's energy loss is described by fragmentation function which is independent of the process. While for heavy mesons the shape of fragmentation function can be extracted from experimental data for $e^+e^-$ annihilation, it remains unknown in case of doubly heavy diquark. However it is reasonable to suppose that this function is quite sharp even for $cc$-diquark because of its relatively high mass.

\section{Production of doubly charmed baryons with excited heavy diquark and prospects for their observation} 

For estimations of cross-sections and yields of doubly charmed baryons in hadron-hadron interactions we have used wave functions from~\cite{Ebert:2002ig} and parton functions CTEQ~\cite{Dulat:2016rzo}.

\begin{figure}[ht]
\begin{tabular}{cc}
\includegraphics[width=0.5\textwidth]{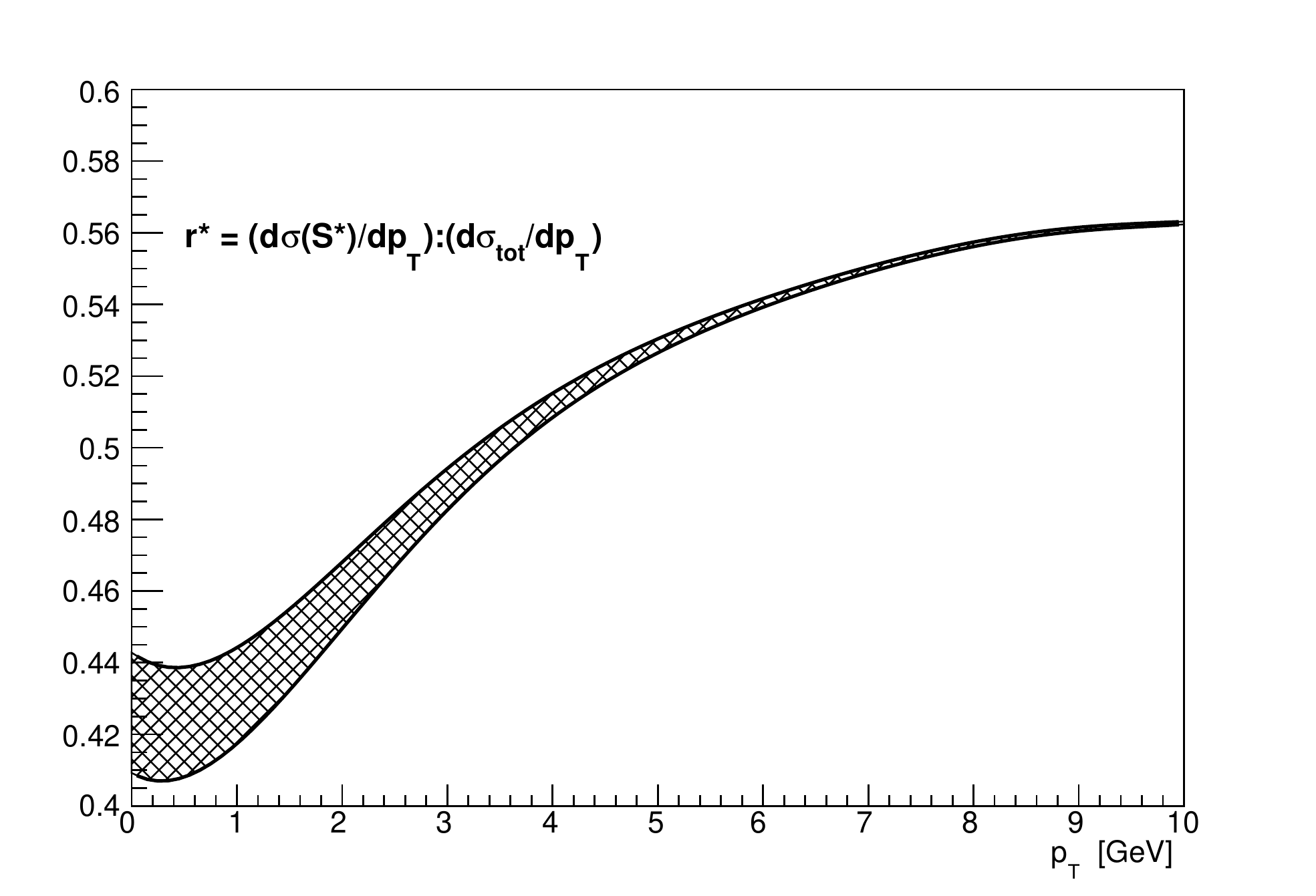} & 
\includegraphics[width=0.5\textwidth]{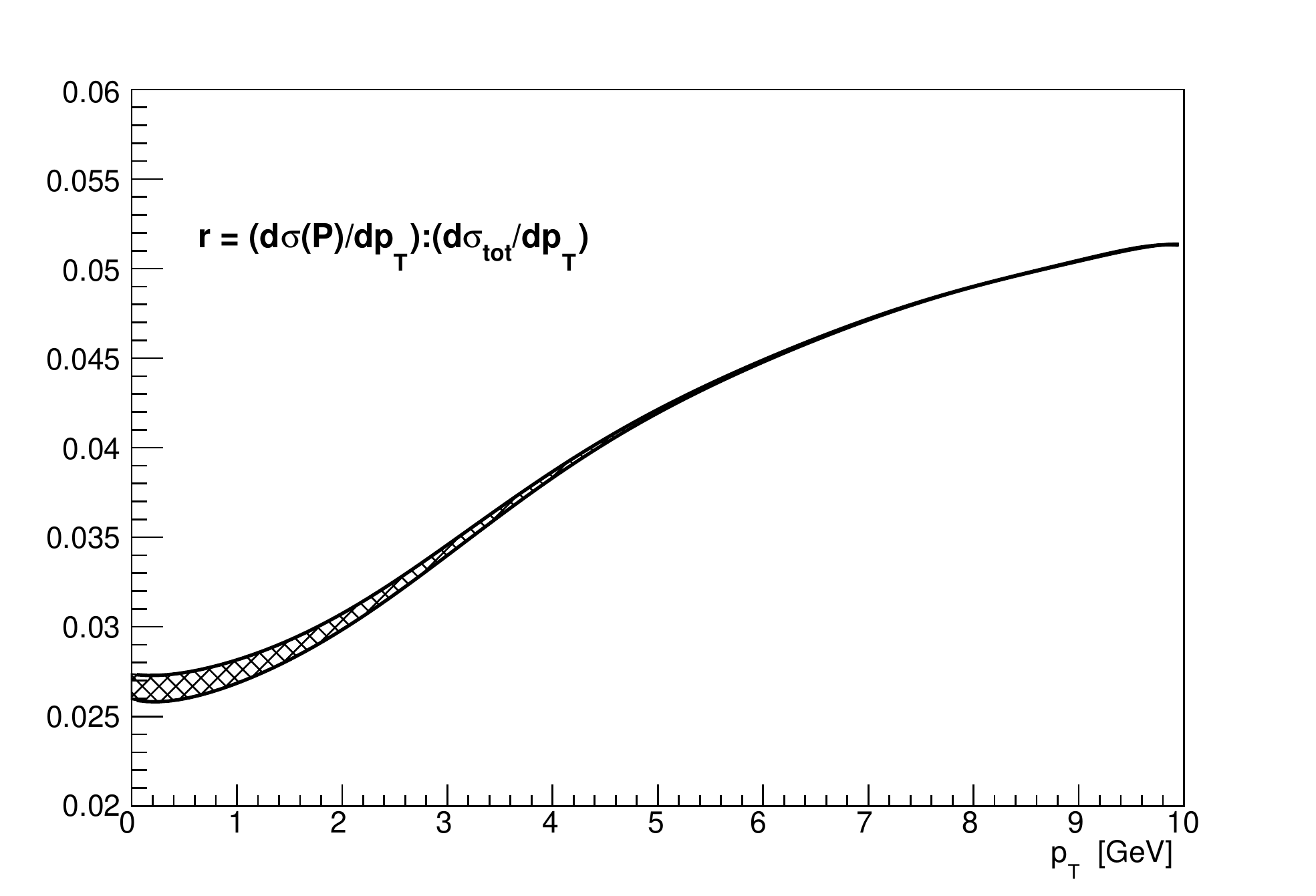} \\
(a) & (b)  
\end{tabular}
\caption{Dependence of relative yields of doubly charmed diquark's excited states on transverse momentum for different scales at proton-proton interaction energy $\sqrt{s}=13$ TeV: (a) -- $S$-wave states, (b)~--~$P$-wave states.}
\label{fig:Xicc}
\end{figure}

The calculations are performed for kinematics of the detector LHCb $\ 2<\eta<4.5,\ p_T<10$ GeV at center-of-mass energy $\sqrt{s}=13$ TeV for scales in the range from $E_T/2$ to $2E_T$.  Our estimations derive that relative yields of baryons with doubly charmed diquark in $2S$- and $3S$-states comprise about 50\%, while $P$-wave states of diquark give only $3\div5$~\% of the total yield (see Tab.~\ref{tab:Xicc} and Fig.~\ref{fig:Xicc}). The estimations obtained show that relative contribution of excited states slightly increases with transverse momentum. However that doesn't mean excited states to be searched at high transverse momenta since absolute yields are higher at low momenta~\cite{Berezhnoy:2018krl}.

\begin{table}[ht]
\caption{Wave functions and masses of doubly charmed diquark~\cite{Ebert:2002ig}. Cross-sections and relative yields for different states of $cc$-diquark.}
\label{parameters_c}
\centering
\begin{tabular}{||c|c|c||c|c||}\hline
state & wave function & diquark's mass & relative yield & cross-section \\
\hline
 & $|R(0)|$, GeV$^{3/2}$ & $m$, GeV & $r^*$,\% & $\sigma$, nb \\
$1S$  & 0.566 &  3.20 & $49\div52$ & $120\div170$\\
$2S$  & 0.540 &  3.50 & $26\div27$ & $60\div90$ \\
$3S$  & 0.542 &  3.70 & $18\div20$ & $40\div70$\\\hline\hline
 & $|R'(0)|$, GeV$^{5/2}$ & $m$, GeV  &  $r$, \% & $\sigma$, nb  \\
$1P$  & 0.149 &  3.40  & 2 & $4\div6$\\
$2P$  &  0.198 &  3.70 & $1\div2$ & $4\div5$\\
\hline
\end{tabular}
\label{tab:Xicc}
\end{table}

Now having gained an estimation for yields of doubly charmed baryons it is worth to discuss their decays.

The excited states of doubly charmed baryons, lying below the $\Lambda_c D$ threshold, fall into the ground state. Where it is kinematically possible the hadronic mode dominates: predictions for widths of electromagnetic transitions~\cite{Dai:2000hza,Lu:2017meb,Xiao:2017udy} at least two orders less than for hadronic ones~\cite{Ma:2017nik,Ma:2015lba,Xiao:2017dly,Xiao:2017udy,Mehen:2017nrh,Hu:2005gf}.

 Since quark-diquark model of doubly heavy baryons allows one to examine separately the excitations of a light quark and a heavy diquark, following this approach one can categorize transitions between the different states of doubly heavy baryons into transitions caused by a change of the light quark state in the baryon and transitions caused by a change of the diquark state.

While $\lambda$-excitations of doubly charmed baryons are predicted to be rather broad by all the theoretic groups~\cite{Ma:2017nik,Ma:2015lba,Xiao:2017dly,Xiao:2017udy}: $40\div300$ MeV, in case of $\rho$-excitations, being studied in the present work, results of different studies contradict each other. Thus according to predictions~\cite{Mehen:2017nrh} the decay widths of doubly charmed baryons with first radial excitation of the diquark and with excitations of light degree of freedom are comparable in magnitude~\footnote{Hereinafter we use generally accepted notations where a number and an uppercase letter denote the heavy diquark orbital state, a number and a lowercase letter -- the light quark orbital state, and a number in parentheses -- the total angular momentum of the baryon.}:
\begin{align*}
&\Gamma\left[\Xi_{cc}\left(2S1s(1/2)\right)\to\Xi_{cc}(1S1s) \right] \sim 50 \text{\ MeV},\\ &\Gamma\left[\Xi_{cc}\left(2S1s(3/2)\right)\to\Xi_{cc}(1S1s)\right] \sim 400 \text{\ MeV},
\end{align*}
which contradicts research~\cite{Eakins:2012fq}, where the values less than $0.5$~MeV are predicted for the transition widths $\Xi_{cc}(2S)\to\Xi_{cc}(1S)\pi$.

The most interesting states in family of doubly charmed baryons are doubly charmed baryons with $P$-wave state of the heavy diquark. The point is, as it was shown in~\cite{Gershtein:1998un}, their decays should be accompanied by a simultaneous change of the spin and orbital momentum of the diquark which leads to width suppression with a factor $\Lambda_{QCD}^2/m_c^2$. Therefore doubly charmed baryons with $P$-wave state of heavy diquark are metastable. This conclusion is partially confirmed by conclusions of research~\cite{Hu:2005gf}, where widths of states $\Xi_{cc}(1P)$ are estimated as follows:
 
\begin{equation}
\begin{cases}
&\Gamma\left[\Xi_{cc}(1P1s(3/2)) \to \Xi_{cc}(1S1s(3/2)) \pi\right] = \lambda_{3/2}^{2} 112\ \text{MeV},\\  
&\Gamma\left[\Xi_{cc}(1P1s(1/2))\to \Xi_{cc}(1S1s(1/2)) \pi \right] = \lambda_{1/2}^{2} 111\ \text{MeV},
\end{cases}
\label{eq:p_Xicc_widths}
\end{equation}
where $\lambda_{3/2},\lambda_{1/2} \sim \Lambda_{QCD}/m_c$. It's clear that at small values of $\lambda_{1/2}$ and $\lambda_{3/2}$ these states will indeed be metastable.
 
\begin{figure}[ht]
\centering
 \includegraphics[width = 0.7\linewidth]{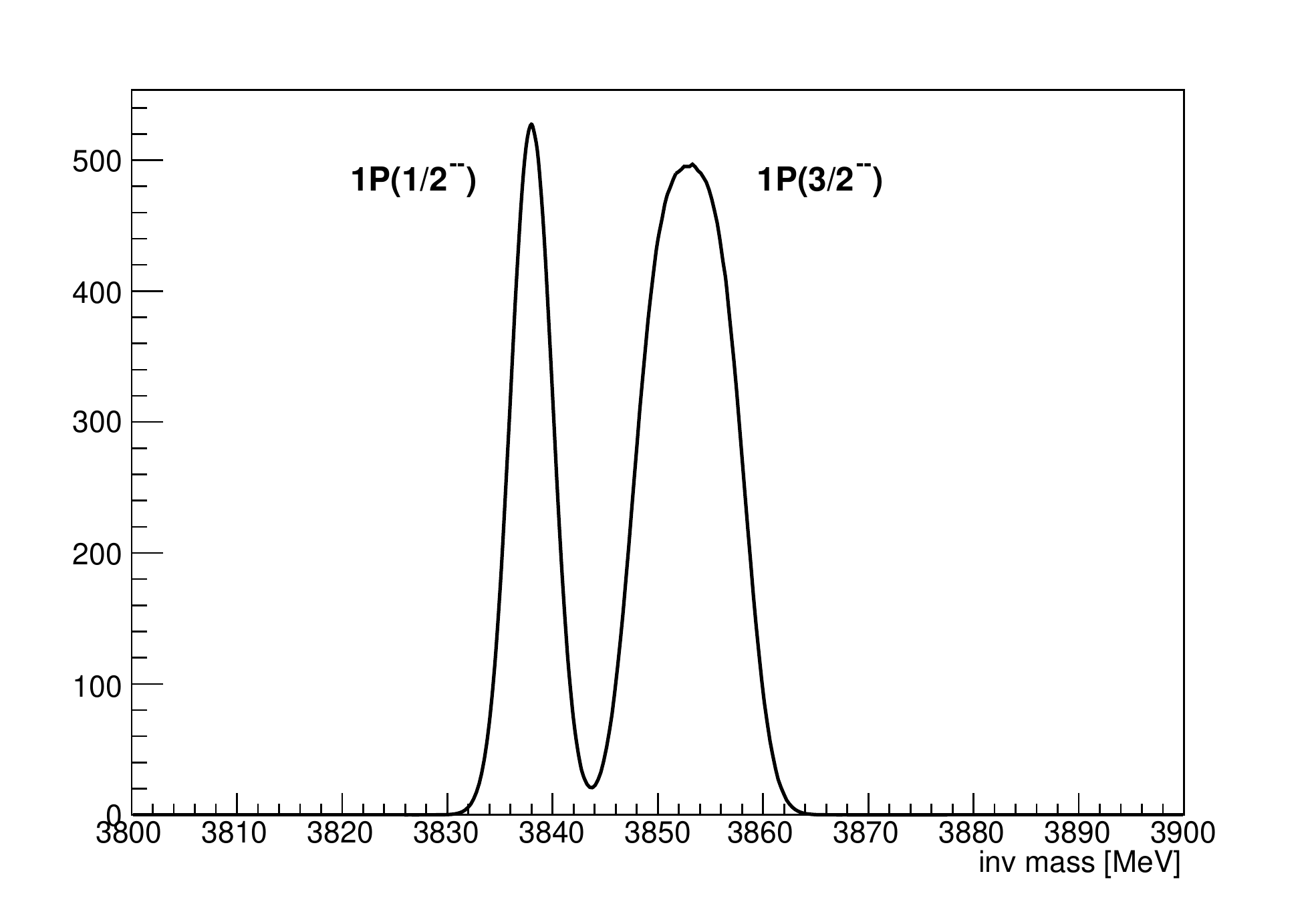}
 \caption{Possible distribution over mass for $1P1s$- levels of $\Xi_{cc}$. Both peaks are washed out with dispersion of $10~\text{MeV}$ in order to simulate the detector's resolution. The $J^P = 3/2^-$ peak’s offset and extra widening are associated with the loss of a soft photon.} 
 \label{fig:peaks1P}
\end{figure}

At LHC experiments conditions these $P$-wave states can be detected through their decays with charge change. 
Decays 
 $\Xi_{cc}^{++}\left(1P1s(1/2)\right) \to \Xi_{cc}^{+}\left(1S1s(1/2)\right)\pi^+$ and $\Xi_{cc}^{+}\left(1P1s(1/2)\right) \to \Xi_{cc}^{++}\left(1S1s(1/2)\right)\pi^-$ can be completely reconstructed. Decays $\Xi_{cc}^{++}\left(1P1s(3/2)\right) \to \Xi_{cc}^{+}\left(1S1s(3/2)\right)\pi^+ \to \left[\Xi_{cc}^{+}\left(1S1s(1/2)\right)\gamma\right]\pi^{+}$ and $\Xi_{cc}^{+}\left(1P1s(3/2)\right) \to \Xi_{cc}^{++}\left(1S1s(3/2)\right)\pi^-\to \left[\Xi_{cc}^{++}\left(1S1s(1/2)\right)\gamma\right]\pi^{-}$  can be reconstructed with photon loss only, since such a soft photon has low detection efficiency. However it will be anyway possible to distinguish the peak, corresponding to $\Xi_{cc}\left(1P1s(3/2)\right)$, in the distribution over $\Xi_{cc}\pi$ invariant mass. This peak will be shifted by the value of mass splitting of doublet $1S1s$ and will have an extra width:
\begin{equation}
\Delta\tilde M \approx 2\Delta M^S\sqrt{\left(\Delta M^{PS}/M\right)^2 - \left(m_{\pi}/M\right)^2 } \sim 10\ \text{MeV},
\end{equation}
where $M$ -- ground state mass, $m_\pi$ -- pion mass,  $\Delta M^S=M\left(\Xi_{cc}(1S1s(3/2)\right) - M\left(\Xi_{cc}(1S1s(1/2)\right)=M\left(\Xi_{cc}(1S1s(3/2)\right) - M$, and $\Delta M^{PS}$ is a mass difference between $1P1s(3/2)$ and $1S1s(3/2)$ states: $\Delta M^{PS}=M\left(\Xi_{cc}(1P1s(3/2)\right) - M\left(\Xi_{cc}(1S1s(3/2)\right)$.  Fig.~\ref{fig:peaks1P} illustrates the possible peaks shape in the invariant mass distribution for candidates to first $P$-wave excitation of the diquark in double charmed baryon. It is worth to add that the transition in $1S1s$ doublet can go  through photon emission only since the value of mass splitting $\Delta M^{S}$ is around~$100\div130~\text{MeV}$~\cite{Ebert:2002ig,Brambilla:2005yk,Fleming:2005pd,Kiselev:2017eic}, that is less than the pion mass.

In $\Omega_{cc}$ spectrum analogous single-pion transitions break the isospin symmetry and therefore, if kinematically possible, the $\Omega_{cc}$ excitations decay into $\Xi_{cc}$ ground state with kaon emission. A special case is the first P-wave diquark excitation in $\Omega_{cc}$. The single-pion transitions are suppresed over three orders due to isospin symmetry breaking~\cite{Ma:2017nik}, and the single-kaon transitions are kinematically forbidden. As a result, for such states the hadronic mode doesn't dominate towards the electromagnetic one (see~\cite{Dai:2000hza} and~\cite{Ma:2017nik}).

It seems that decay widths of $2S$-states exceed the value $\Delta M^S$ of hyperfine splitting and therefore in case of transition $\Xi_{cc}\left(2S\right) \to \Xi_{cc}\left(1S\right)\pi$ the quantum numbers $J^P$ are unlikey to be determined: decay will be represented by one wide peak in the invariant mass distribution.

Basing on the estimations concerning the relative yield of excited baryons, we can roughly conclude how many of doubly charmed baryons represent the decay products of excited states $\Xi_{cc}^*$ and $\Omega_{cc}^*$. The total yield of $\Xi_{cc}^{++}$, seen in experiment, might be estimated as
\begin{multline}\notag
N_{tot} \sim N_{direct}(\Xi_{cc}^{++}) + \frac{1}{3}N(\Xi_{cc}^{*++} \to \Xi_{cc}^{++}\pi^0) + \\ + \frac{2}{3}N(\Xi_{cc}^{*+} \to \Xi_{cc}^{++}\pi^-) + \frac{1}{2} \cdot \frac{1}{2} N(\Omega_{cc}^{*+} \to \Xi_{cc}^{++}K^-).
\end{multline}

Here the coefficients at $N(\Xi_{cc}^{*++} \to \Xi_{cc}^{++}\pi^0)$ and $N(\Xi_{cc}^{*+} \to \Xi_{cc}^{++}\pi^-)$ are determined by isospin count and coefficient at $N(\Omega_{cc}^{*+} \to \Xi_{cc}^{++}K^-)$ is determined by isospin count and consideration that about half of excited states of $\Omega_{cc}^{*+}$ might lie under the threshold of $\Xi_{cc}^{++}K^-$. Assuming that excited diquarks hadronize by picking up the light quark in the same proportion as not excited ones $ u: d: s = 1:1:0.26 $, one can conclude that from $N_{tot} \approx 300$  detected $\Xi_{cc}^{++}$ around $\frac{2/3}{2.26}\times 300 \approx 90$ baryons are products of $\Xi_{cc}^{*+}$ decay, around $\frac{2/3}{2.26}\times 300 \approx 45$ baryons are products of $\Xi_{cc}^{*++}$ decay and around $\frac{0.26/4}{2.26}\times 300 \approx 10$ baryons are decay products of $\Omega_{cc}^{*+}$.
 
\section{Excited doubly beauty baryons and prospects for their observation}

It is worth noting that even perspectives for ground state observation are not clear so far because of the very low production cross-section. It is necessary to produce four beauty quarks to form such a state, which results in high suppression because of the small phase space at low gluon energies and to suppression as $|R(0)|^2/m_b^3$ to compare with $b$ quarks production at high energies in the fragmentation regime. Most probably there is no enhancement caused by double parton scattering in these processes. Nevertheless the possibility of searching for these states is being discussed. So we should mention a very interesting paper~\cite{Gershon:2018gda}, where the possibility of discovering $\Xi_{bb}$ is regarded by means of detection of $B_c$ mesons with momentum directed not to primary interaction vertex and therefore with high probability representing the $\Xi_{bb}$ decay products. Considering a certain attention to the problem of $\Xi_{bb}$ registration, in this work we estimate the relative yields of $\Xi_{bb}$ with $S$- and $P$-wave excitations of diquark.

\begin{figure}[ht]
\begin{tabular}{cc}
\includegraphics[width=0.5\textwidth]{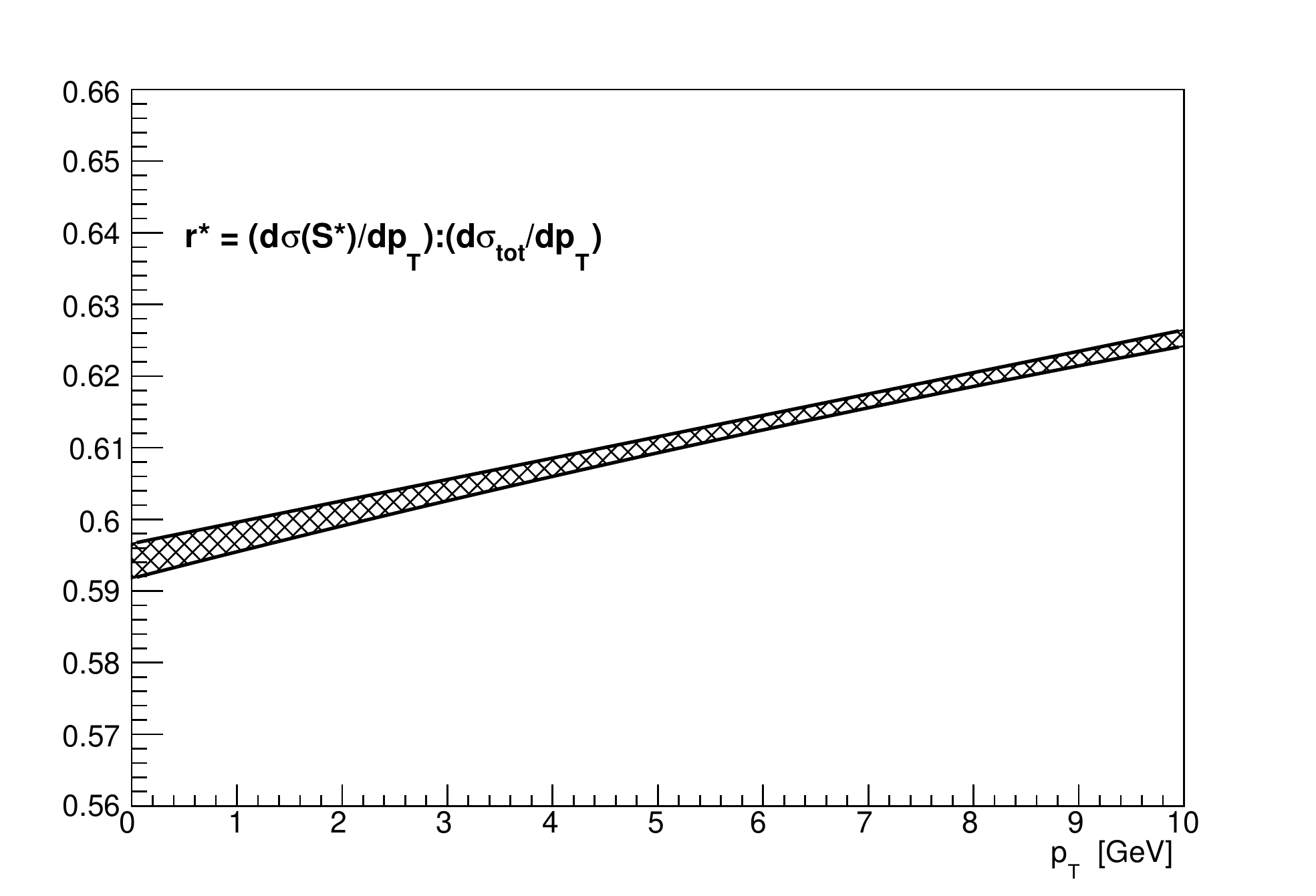} & 
\includegraphics[width=0.5\textwidth]{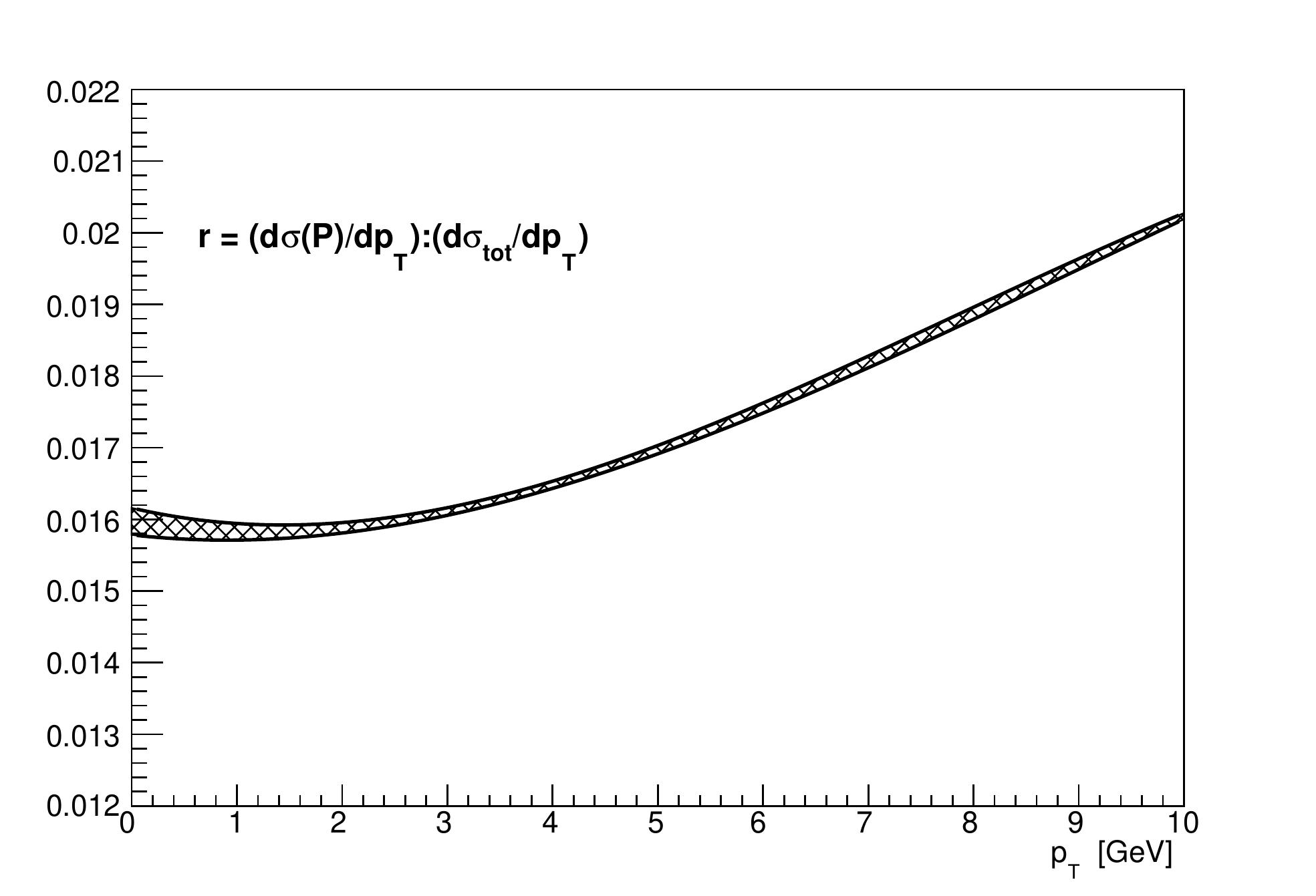} \\
(a) & (b)  
\end{tabular}
\caption{Dependence of relative yields of doubly beauty diquark's excited states on transverse momentum for different scales at proton-proton interaction energy $\sqrt{s}=13$ TeV: (a) -- $S$-wave states, (b)~--~$P$-wave states.}
\label{fig:Xibb}
\end{figure}

Using results for masses and wave functions gained in papers~\cite{Ebert:2002ig,Galkin:2020Xibb}, we perform our estimations of cross-sections and relative yields for excited states (see Fig.~\ref{fig:Xibb}). They have revealed that yield of metastable $P$-wave states of doubly beauty baryons is suppressed even more than in case of doubly charmed baryons and is approximately 2\% from the total yield of all the doubly beauty baryons. Meanwhile contribution from $S$-wave states, which is about 60\%, is a little more, than contribution from analogous states to the yield of doubly charmed baryons, which is $\sim 50$\%.
 
\begin{table}[ht]
\caption{Wave functions and masses of doubly beauty diquark~\cite{Ebert:2002ig,Galkin:2020Xibb}. Cross-sections and relative yields for different states of $bb$-diquark.}
\label{parameters_b}
\centering
\begin{tabular}{||c|c|c||c|c||}\hline
state & wave function & diquark's mass & relative yield & cross-section\\
\hline
 & $|R(0)|$, GeV$^{3/2}$ & $m$, GeV & $r^*$,\% & $\sigma$, pb \\
$1S$  &  1.107&  9.8 & $36\div37$ & $320\div670$  \\
$2S$  &  0.969&  10.0 & $24\div25$  & $210\div450$ \\
$3S$  &  0.927 &  10.2 & $19\div20$ & $170\div360$ \\
$4S$  &  0.906 &  10.3 & $17\div18$ & $150\div320$ \\\hline\hline
 & $|R'(0)|$, GeV$^{5/2}$ & $m$, GeV  &  $r$, \% & $\sigma$, pb  \\
$1P$  &  0.387 &  9.9  & 0.3 & $3\div6$\\
$2P$  &  0.484 &  10.1 & 0.4 & $4\div8$\\
$3P$  &  0.551 &  10.3 & 0.5 & $4\div9$\\
$4P$  &  0.605 &  10.4 & 0.5 & $4\div9$\\
\hline
\end{tabular}
\label{tab:Xibb}
\end{table}

Since $bb$-diquark is more compact, most likely the quark-diquark model should describe the family of doubly beauty baryons more successfully than the family of doubly charmed baryons because of the smaller corrections on the diquark's size~\cite{Gershtein:2000nx}. It is also notable that metastable baryons with scalar $P$-wave $bb$-diquark should be even more narrow than analogous states with $cc$-diquark: their  widths should be approximately in $m_b^2/m_c^2$ times less as evidenced by equation~(\ref{eq:p_Xicc_widths}).

\section{Conclusion}

In this study the relative yields of doubly charmed and doubly beauty baryons with excited heavy diquark are estimated. In both cases the excitations with $S$-wave diquark's state account for about half from the total yield of such double heavy baryons and excitations with $P$-wave diquark's state account for only few percent from the total yield. The calculations show that search for $S$-wave excitations of doubly charmed baryons is quite feasible challenge for LHCb experiment. Searching for $P$-wave states of doubly charmed baryons represents much more complicated challenge. Detection perspectives for excitations of $\Xi_{bb}$ baryons in the experiments at LHC at this point remain in doubt.

This research was done with support of RFBR grant № 20-02-00154~A. The work of A.V.~Berezhony and I.N.~Belov is supported by ``Basis'' Foundation, grants №№~17-12-244-1  and 7-12-244-41.

\vspace{5ex} 
\bibliographystyle{unsrt}
\bibliography{all} 

\begin{thebibliography}{10}

\bibitem{Ebert:1996ec}
D.~Ebert, R.~N. Faustov, V.~O. Galkin, A.~P. Martynenko, and V.~A. Saleev.
\newblock {Heavy baryons in the relativistic quark model}.
\newblock {\em Z. Phys.}, C76:111--115, 1997.

\bibitem{Kiselev:2001fw}
V.~V. Kiselev and A.~K. Likhoded.
\newblock {Baryons with two heavy quarks}.
\newblock {\em Phys. Usp.}, 45:455--506, 2002.
\newblock [Usp. Fiz. Nauk172,497(2002)].

\bibitem{Kiselev:2002an}
V.V. Kiselev and A.K. Likhoded.
\newblock {Comment on `First observation of doubly charmed baryon Xi(cc)+'}.
\newblock 2002.

\bibitem{Faustov:2018vgl}
Rudolf~N. Faustov and Vladimir~O. Galkin.
\newblock {Heavy baryon spectroscopy}.
\newblock {\em EPJ Web Conf.}, 204:08001, 2019.

\bibitem{Kerbikov:1987vx}
B.~O. Kerbikov, M.~I. Polikarpov, and L.~V. Shevchenko.
\newblock {Multi - Quark Masses and Wave Functions Through Modified Green's
  Function Monte Carlo Method}.
\newblock {\em Nucl. Phys.}, B331:19, 1990.

\bibitem{Albertus:2006wb}
C.~Albertus, E.~Hernandez, J.~Nieves, and J.~M. Verde-Velasco.
\newblock {Doubly heavy quark baryon spectroscopy and semileptonic decay}.
\newblock {\em Eur. Phys. J.}, A31:691--694, 2007.

\bibitem{Albertus:2006ya}
C.~Albertus, E.~Hernandez, J.~Nieves, and J.~M. Verde-Velasco.
\newblock {Static properties and semileptonic decays of doubly heavy baryons in
  a nonrelativistic quark model}.
\newblock {\em Eur. Phys. J.}, A32:183--199, 2007.
\newblock [Erratum: Eur. Phys. J.A36,119(2008)].

\bibitem{Roncaglia:1995az}
R.~Roncaglia, D.~B. Lichtenberg, and E.~Predazzi.
\newblock {Predicting the masses of baryons containing one or two heavy
  quarks}.
\newblock {\em Phys. Rev.}, D52:1722--1725, 1995.

\bibitem{Roberts:2007ni}
W.~Roberts and Muslema Pervin.
\newblock {Heavy baryons in a quark model}.
\newblock {\em Int. J. Mod. Phys.}, A23:2817--2860, 2008.

\bibitem{Yoshida:2015tia}
Tetsuya Yoshida, Emiko Hiyama, Atsushi Hosaka, Makoto Oka, and Katsunori
  Sadato.
\newblock {Spectrum of heavy baryons in the quark model}.
\newblock {\em Phys. Rev.}, D92(11):114029, 2015.

\bibitem{Kom:2011bd}
C.H. Kom, A.~Kulesza, and W.J. Stirling.
\newblock {Pair production of J/psi as a probe of double parton scattering at
  LHCb}.
\newblock {\em Phys.Rev.Lett.}, 107:082002, 2011.

\bibitem{Baranov:2011ch}
S.P. Baranov, A.M. Snigirev, and N.P. Zotov.
\newblock {Double heavy meson production through double parton scattering in
  hadronic collisions}.
\newblock {\em Phys.Lett.}, B705:116--119, 2011.

\bibitem{Berezhnoy:2012xq}
A.V. Berezhnoy, A.K. Likhoded, A.V. Luchinsky, and A.A. Novoselov.
\newblock {Associated production of $J/\psi$-mesons and open charm and double
  open charm production at the LHC}.
\newblock 2012.

\bibitem{Berezhnoy:2015jga}
A.V. Berezhnoy and A.K. Likhoded.
\newblock {Associated production of Y and open charm at LHC}.
\newblock {\em Int. J. Mod. Phys. A}, 30(20):1550125, 2015.

\bibitem{Aaij:2017ueg}
Roel Aaij et~al.
\newblock {Observation of the doubly charmed baryon $\Xi_{cc}^{++}$}.
\newblock {\em Phys. Rev. Lett.}, 119(11):112001, 2017.

\bibitem{Aaij:2018gfl}
Roel Aaij et~al.
\newblock {First observation of the doubly charmed baryon decay
  $\Xi_{cc}^{++}\rightarrow \Xi_{c}^{+}\pi^{+}$}.
\newblock {\em Phys. Rev. Lett.}, 121:162002, 2018.

\bibitem{Aaij:2018wzf}
Roel Aaij et~al.
\newblock {First measurement of the lifetime of the doubly charmed baryon
  Xi(cc)++}.
\newblock {\em Phys. Rev.Lett}, 121:052002, 2018.

\bibitem{Berezhnoy:2018krl}
A.~V. Berezhnoy, I.~N. Belov, and A.~K. Likhoded.
\newblock {Production of doubly charmed baryons with the excited heavy diquark
  at LHC}.
\newblock {\em Int. J. Mod. Phys.}, A34(06n07):1950038, 2019.

\bibitem{Berezhnoy:1995fy}
A.~V. Berezhnoy, V.~V. Kiselev, and A.~K. Likhoded.
\newblock {Hadronic production of baryons containing two heavy quarks}.
\newblock {\em Phys. Atom. Nucl.}, 59:870--874, 1996.
\newblock [Yad. Fiz.59,909(1996)].

\bibitem{Berezhnoy:1998aa}
A.~V. Berezhnoy, V.~V. Kiselev, A.~K. Likhoded, and A.~I. Onishchenko.
\newblock {Doubly charmed baryon production in hadronic experiments}.
\newblock {\em Phys. Rev.}, D57:4385--4392, 1998.

\bibitem{Baranov:1995rc}
S.~P. Baranov.
\newblock {On the production of doubly flavored baryons in p p, e p and gamma
  gamma collisions}.
\newblock {\em Phys. Rev.}, D54:3228--3236, 1996.

\bibitem{Baranov:1997sg}
S.~P. Baranov.
\newblock {Semiperturbative and nonperturbative production of hadrons with two
  heavy flavors}.
\newblock {\em Phys. Rev.}, D56:3046--3056, 1997.

\bibitem{Chang:2006eu}
Chao-Hsi Chang, Cong-Feng Qiao, Jian-Xiong Wang, and Xing-Gang Wu.
\newblock {Estimate of the hadronic production of the doubly charmed baryon
  Xi(cc) under GM-VFN scheme}.
\newblock {\em Phys. Rev.}, D73:094022, 2006.

\bibitem{Gershtein:2000nx}
S.~S. Gershtein, V.~V. Kiselev, A.~K. Likhoded, and A.~I. Onishchenko.
\newblock {Spectroscopy of doubly heavy baryons}.
\newblock {\em Phys. Rev. D}, 62:054021, 2000.

\bibitem{Aaij:2011jp}
R.~Aaij et~al.
\newblock {Measurement of $b$-hadron production fractions in $7~\rm{TeV} pp$
  collisions}.
\newblock {\em Phys.Rev.}, D85:032008, 2012.

\bibitem{Ebert:2002ig}
D.~Ebert, R.~N. Faustov, V.~O. Galkin, and A.~P. Martynenko.
\newblock Mass spectra of doubly heavy baryons in the relativistic quark model.
\newblock {\em Phys. Rev.}, D66:014008, 2002.

\bibitem{Dulat:2016rzo}
S.~Dulat, T.~J. Hou, J.~Gao, M.~Guzzi, J.~Huston, P.~Nadolsky, J.~Pumplin,
  C.~Schmidt, D.~Stump, and C.~P. Yuan.
\newblock {The structure of the proton: The CT14 QCD global analysis}.
\newblock {\em EPJ Web Conf.}, 120:07003, 2016.

\bibitem{Dai:2000hza}
Wu-Sheng Dai, Xin-Heng Guo, Hong-Ying Jin, and Xue-Qian Li.
\newblock {Electromagnetic radiation of baryons containing two heavy quarks}.
\newblock {\em Phys. Rev.}, D62:114026, 2000.

\bibitem{Lu:2017meb}
Qi-Fang Lu, Kai-Lei Wang, Li-Ye Xiao, and Xian-Hui Zhong.
\newblock {Mass spectra and radiative transitions of doubly heavy baryons in a
  relativized quark model}.
\newblock {\em Phys. Rev.}, D96(11):114006, 2017.

\bibitem{Xiao:2017udy}
Li-Ye Xiao, Kai-Lei Wang, Qi-fang Lu, Xian-Hui Zhong, and Shi-Lin Zhu.
\newblock {Strong and radiative decays of the doubly charmed baryons}.
\newblock {\em Phys. Rev.}, D96(9):094005, 2017.

\bibitem{Ma:2017nik}
Yong-Liang Ma and Masayasu Harada.
\newblock {Chiral partner structure of doubly heavy baryons with heavy quark
  spin-flavor symmetry}.
\newblock 2017.

\bibitem{Ma:2015lba}
Yong-Liang Ma and Masayasu Harada.
\newblock {Doubly heavy baryons with chiral partner structure}.
\newblock {\em Phys. Lett.}, B748:463--466, 2015.

\bibitem{Xiao:2017dly}
Li-Ye Xiao, Qi-Fang Lü, and Shi-Lin Zhu.
\newblock {Strong decays of the 1P and 2D doubly charmed states}.
\newblock {\em Phys. Rev.}, D97(7):074005, 2018.

\bibitem{Mehen:2017nrh}
Thomas Mehen.
\newblock {Implications of Heavy Quark-Diquark Symmetry for Excited Doubly
  Heavy Baryons and Tetraquarks}.
\newblock {\em Phys. Rev.}, D96(9):094028, 2017.

\bibitem{Hu:2005gf}
Jie Hu and Thomas Mehen.
\newblock {Chiral Lagrangian with heavy quark-diquark symmetry}.
\newblock {\em Phys. Rev.}, D73:054003, 2006.

\bibitem{Eakins:2012fq}
B.~Eakins and W.~Roberts.
\newblock {Heavy Diquark Symmetry Constraints for Strong Decays}.
\newblock {\em Int. J. Mod. Phys.}, A27:1250153, 2012.

\bibitem{Gershtein:1998un}
S.~S. Gershtein, V.~V. Kiselev, A.~K. Likhoded, and A.~I. Onishchenko.
\newblock {Spectroscopy of doubly heavy baryons}.
\newblock {\em Acta Phys. Hung.}, A9:133--144, 1999.
\newblock [Yad. Fiz.63,334(2000)].

\bibitem{Brambilla:2005yk}
Nora Brambilla, Antonio Vairo, and Thomas Rosch.
\newblock {Effective field theory Lagrangians for baryons with two and three
  heavy quarks}.
\newblock {\em Phys.Rev.}, D72:034021, 2005.

\bibitem{Fleming:2005pd}
Sean Fleming and Thomas Mehen.
\newblock {Doubly heavy baryons, heavy quark-diquark symmetry and NRQCD}.
\newblock {\em Phys. Rev.}, D73:034502, 2006.

\bibitem{Kiselev:2017eic}
V.~V. Kiselev, A.~V. Berezhnoy, and A.~K. Likhoded.
\newblock {Quark–Diquark Structure and Masses of Doubly Charmed Baryons}.
\newblock {\em Phys. Atom. Nucl.}, 81(3):369--372, 2018.
\newblock [Yad. Fiz.81,no.3,356(2018)].

\bibitem{Gershon:2018gda}
T.~Gershon and A.~Poluektov.
\newblock Displaced $b_c^-$ mesons as an inclusive signature of weakly decaying
  double beauty hadrons.
\newblock {\em JHEP}, 01:019, 2019.

\bibitem{Galkin:2020Xibb}
V.O. Galkin.
\newblock Private communications.
\newblock Wave functions and masses for $\Xi_{QQ}$ states, 2020.

\end{thebibliography}
\end{document}